# AC electrokinetic phenomena over semiconductive surfaces: effective electric boundary conditions and their applications

Cunlu Zhao[*] and Chun Yang

*School of Mechanical and Aerospace Engineering, Nanyang Technological University*

*50 Nanyang Avenue, Singapore 639798, Republic of Singapore*

* To whom correspondence should be addressed. E-mail: zhao0070@e.ntu.edu.sg

## Abstract

Electrokinetic boundary conditions are derived for AC electrokinetic (ACEK) phenomena over leaky dielectric (i.e., semiconducting) surfaces. Such boundary conditions correlate the electric potentials across the semiconductor-electrolyte interface (consisting of the electric double layer (EDL) inside the electrolyte solutions and the space charge layer (SCL) inside the semiconductors) under AC electric fields with arbitrary wave forms. The present electrokinetic boundary conditions allow for evaluation of induced zeta potential contributed by both bond charges (due to electric polarization) and free charges (due to electric conduction) from the leaky dielectric materials. Subsequently, we demonstrate the applications of these boundary conditions in analyzing the ACEK phenomena around a semiconducting cylinder. It is concluded that the flow circulations exist around the semiconducting cylinder and are shown to be stronger under an AC field with lower frequency and around a cylinder with higher conductivity.

## I. INTRODUCTION

AC electrokinetic (ACEK) phenomena are widely used for manipulations of particles and flows in microfluidic systems [1-3]. Classic description of electrokinetic phenomena relies on the electric double layer (EDL) formed on charged insulating surfaces whose surface charge density are fixed due to the physiochemical bonds. Consequently, the surface charge density is independent of applied electric fields. However, for electrokinetic phenomena happens around polarizable or conducting solids, it has been confirmed that in the presence of an external electric field, extra electric charges can be induced on polarizable or conducting solid surfaces immersed in electrolyte solutions, thereby triggering the charging of EDL inside the liquid. This is manifested in a zeta potential which is no longer a fixed equilibrium material property, but rather depends upon the externally applied electric field. These induced-charge electrokinetic (ICEK) phenomena have been studied firstly for polarizable colloidal particles [4, 5] almost two decades ago and recently in the context of microfluidic applications for pumping [6, 7], mixing [6, 8, 9], demixing [10], focusing [11] and particle manipulations [12, 13].

Until recently, attention has been however mainly focused on ICEK phenomena of conductors with ideal polarizability. Investigations of electrokinetic phenomena over dielectric surfaces with finite polarizability just emerged, and the induced zeta potential in this case is solely contributed by the bond charges due to polarization. Squires and Bazant [6] analyzed a decrease in the induced zeta potential due to the presence of a thin dielectric coating on a conducting surface. For solids with arbitrary polarizability, two same effective electric boundary conditions of a Robin-type and a Neumann-type are derived via different methodologies in Refs. [14, 15] for evaluations of induced zeta potentials on arbitrarily polarizable dielectric surfaces. Their analyses both confirmed that dielectric surfaces with stronger polarizability acquire higher induced zeta potentials and thus induce stronger electrokinetic phenomena. Yossifon et al [16] further derived the transient version of these effective boundary conditions, which allow for predicting transient development of the induced zeta potential over arbitrarily polarizable dielectric surfaces. Using these derived electric boundary conditions, they investigated the temporary evolution of DC driven electrokinetic phenomena around a polarizable object. Only very recently, Pascall and squires [17] experimentally verified that the very presence (often inevitable) of a thin dielectric layer on electrode surfaces (e.g., due to surface contamination by oxidized or adsorbed species) can substantially alter the induced zeta potential, and thus affect the associated electrokinetic phenomena.

However, the aforementioned all assumed that the solids are perfectly dielectric and thus there are no free charge carriers and space charge layer (SCL) inside them. However, for a more general solid, it has both finite dielectric constant and conductivity and is leaky dielectric or semiconductive in nature. Under this case, the SCL forms in the solid and the EDL forms in the liquid, and these two layers constitute the interface between a semiconductor and an electrolyte solution. On the other hand, AC electric forcing is usually much desired in microfluidic applications since it introduces another control parameter of frequency and reduces possible negative effects, such as electrolysis and dissolution of electrodes. Our effort here is to consider both EDL and SCL and derive effective electric boundary conditions eligible to evaluate induced zeta potential on surfaces of semiconducting solids subjected to AC electric field of arbitrary wave forms. Furthermore, we analyzed the ACEK phenomena around a semiconducting cylinder to show the applications of the derived boundary conditions. Finally, possible applications of the derived boundary conditions are also discussed and highlighted.

## II. EFFECTIVE ELECTRIC BOUNDARY CONDITIONS

This section presents electrokinetic boundary conditions for ACEK phenomena to correlate the electrical potentials across the EDL and SCL at a semiconductor-electrolyte solution interface. It is known that there is usually a SCL (i.e., an "EDL" in the solid) in the semiconducting solid adjacent to the EDL in the liquid electrolyte [18-21]. The thicknesses of EDL and SCL have the same order of magnitude and typically range from 1nm to 100nm, thus both of them need to be considered simultaneously. Then we usually need to consider the transport (by diffusion and migration) of both types of charge carriers (i.e., electrons and holes) in the solid as what we usually model both cations and anions in the liquid. We restrict our analysis under the three assumptions (i) thin EDL and SCL, (ii) negligible Peclet number and (iii) weak applied electric field (i.e. $\Psi = ze\Phi_0/(k_B T) \ll 1$, wherein $z$ denotes the valence of charge carriers inside the electrolyte solution, $e$ the elemental charge, $k_B$ the Boltzmann constant and $T$ the absolute temperature ). These three assumptions were also made in previous studies [6, 16, 22]. The thin EDL and SCL assumption requires that the EDL thickness and the SCL thickness are much smaller than the characteristic dimension of the semiconductive solid wall, $a$, that is $\delta_1 = \lambda_{D1}/a \ll 1, \delta_2 = \lambda_{D2}/a \ll 1$. The EDL thickness for a symmetric electrolyte ($z$:$z$) can be defined as $\lambda_{D1} = \sqrt{\varepsilon_0 \varepsilon_f k_B T/(2n_{01}z^2 e^2)}$, and the SCL thickness is similarly defined as $\lambda_{D2} = \sqrt{\varepsilon_0 \varepsilon_w k_B T/(2n_{02} e^2)}$, where $n_{0i}$ denotes the bulk concentration of charge carriers inside the liquid and solid domains ($i$=1 for the electrolytic solution and $i$=2 for the semiconductor, this convention is complied with in the whole work), $\varepsilon_0$ is the electric permittivity of vacuum. $\varepsilon_f$ and $\varepsilon_w$ are the dielectric constants of the electrolyte and semiconductor, respectively.

From a microfluidic application viewpoint, the aforementioned three assumptions are applicable. Therefore, in the model development, the hydrodynamic problem and the electrostatic problem can be decoupled [6, 16, 22]. Furthermore, as shown in Fig. 1, the domain of the electrostatic problem can be divided into four sub-domains: the two domains for the electroneutral bulk solid wall and bulk solution with their harmonic dimensionless electric potentials $\Phi_w$ and $\Phi_f$, respectively, and the two 'inner' domains of the EDL and SCL whose dimensionless electric potentials, $\Phi_{EDL}$ and $\Phi_{SCL}$, satisfy Poisson's equation. Note that all these dimensionless potentials are scaled with respect to the reference potential $\Phi_0$. To obtain the requisite electrokinetic conditions connecting $\Phi_w$ and $\Phi_f$ on the dielectric solid–electrolyte interface, we focus on the inner domains which (for $\delta_1 \ll 1$ and $\delta_2 \ll 1$) is locally one-dimensional in the direction of the $y'$ axis, namely outward normal to the solid surface. We define the corresponding 'outer' ($y$) and 'inner' ($Y_1$, $Y_2$) dimensionless spatial variables through $y' = ay = \lambda_{D1} Y_1 = \lambda_{D2} Y_2$.

In our analysis, the dimensionless net charge densities inside the liquid and solid domains due to the difference of concentrations of positive charge carriers and negative charge carriers, i.e., $n_p - n_n$, is expressed as $\rho_i = \left(n_{pi} - n_{ni}\right)/\left(2\Psi n_{0i}\right)$. To the leading order (in the limit of small $\delta_1$, $\delta_2$ and $\Psi$), $\Phi_{EDL}$ and $\Phi_{SCL}$ respectively satisfy Poisson's equation

$$\frac{\partial^2 \Phi_{EDL}}{\partial Y_1^2} = -\rho_1 \quad \text{and} \quad \frac{\partial^2 \Phi_{SCL}}{\partial Y_2^2} = -z\rho_2 \tag{1}$$

and the continuity equations for electric current (that is obtained from the Nernst–Planck equations for positive and negative carriers in both fluid and solid domains)

$$\frac{\partial \rho_1}{\partial \tau} = \frac{\partial^2 \rho_1}{\partial Y_1^2} - \rho_1 \quad \text{and} \quad \frac{t_w}{t_f}\frac{\partial \rho_2}{\partial \tau} = \frac{\partial^2 \rho_2}{\partial Y_2^2} - \rho_2 \qquad (2)$$

where $\tau$ is the dimensionless time normalized with the reference time $t_f = \lambda_{D1}^2 / D_f$ and in Eq.(2) $t_w = \lambda_{D2}^2 / D_w$ (here $D_f$ stands for the mass diffusivity for free charge carriers in liquid, for a dilute symmetric binary electrolyte it is usually assumed that positive and negative free charge carriers have the same diffusivity, namely $D_{p1} = D_{n1} = D_f$, in the solid wall we also assume that both charge carriers have the save diffusivities, i.e., $D_{p2} = D_{n2} = D_w$). Such reference time $t_f$ also can be expressed as $t_f = \varepsilon_0 \varepsilon_f / \sigma_f = 1/\omega_D$, which denotes the charge relaxation time in the electrolytic solution, and also can be viewed as the time that ions take to travel a Debye length by diffusion. $\omega_D$ is the Debye frequency of the electrolytic solution [23], and $\sigma_f$ is the bulk electric conductivity of the electrolytic solution and can be formulated as $\sigma_f = 2n_{01}z^2e^2 D_f / (k_B T)$. Similarly, the charge relaxation time $t_w = \lambda_{D2}^2 / D_w$ inside the semiconducting solid wall has the same physical interpretation as $t_f$.

On the solid surface, $\Phi_{EDL}$ and $\Phi_{SCL}$ satisfy the electrostatic boundary conditions[24]

$$\Phi_{EDL} = \Phi_{SCL} \quad \text{and} \quad \frac{\partial \Phi_{EDL}}{\partial Y_1} - \beta \frac{\partial \Phi_{SCL}}{\partial Y_2} = -q \quad \text{at } Y_1 \text{ or } Y_2 = 0 \qquad (3)$$

which respectively describe the continuity of the electric potential and the discontinuity of the electric displacement due to the presence of free charges at the interface between two different media. In Eqs. (3), $\beta = (\varepsilon_w \lambda_{D1})/(\varepsilon_f \lambda_{D2})$ and $q$ is the dimensionless free surface charge density which is scaled by $\varepsilon_0 \varepsilon_f \Phi_0 / \lambda_{D1}$.

Under the widely adopted assumption, namely the solid surface is totally blocking and there is no Faradaic reaction on the semiconductive surface, viz., the solid wall is not penetrable to the charge carrier fluxes, the vanishing of the normal components of the positive charge carriers and negative carrier inside both domains leads to the boundary conditions for $\rho_1$ and $\rho_2$ as

$$\frac{\partial \rho_1}{\partial Y_1} = -\frac{\partial \Phi_{EDL}}{\partial Y_1} \qquad \text{at } Y_1=0 \qquad (4a)$$

$$z\frac{\partial \rho_2}{\partial Y_2} = -\frac{\partial \Phi_{SCL}}{\partial Y_2} \qquad \text{at } Y_2=0 \qquad (4b)$$

At the outer edges of the EDL and the SCL, we impose the asymptotic matching conditions as

$$\Phi_{EDL}\big|_{Y_1 \to \infty} = \Phi_f\big|_{y \to 0} \quad \text{and} \quad \Phi_{SCL}\big|_{Y_2 \to -\infty} = \Phi_w\big|_{y \to 0} \qquad (5)$$

and the electroneutrality condition as

$$\rho_1 \to 0 \text{ as } Y_1 \to \infty \qquad (6a)$$

$$\rho_2 \to 0 \text{ as } Y_2 \to -\infty \qquad (6b)$$

We consider time periodic electrokinetic phenomena under an externally applied AC electric field with arbitrary wave form, e.g., sinusoidal, triangular, rectangular, etc. The general time-dependent electric field is assumed to be continuous and to have a piecewise continuous first-order derivative over the period, such that its value at $\tau = \tau + kT_0$ is identical for any integer $k$. Thus, a general periodic field quantity, $X$, can be expressed as a complex Fourier series, $X(\tau) = \sum_{k=-\infty}^{+\infty} X^{(k)} \exp(jk\Omega\tau)$, where $j=\sqrt{-1}$, $\Omega$ is the normalized frequency with respect to the Debye frequency $\omega_D$ of electrolyte solution and it can be computed from $\Omega = 2\pi / T_0$. Here, $X^{(k)}$ represents the complex amplitude of the ambient field such that $X^{(-k)}$ denotes the complex conjugate of $X^{(k)}$. Thus, the aforementioned sum always renders a real function, and $X^{(k)}$ can be defined as $X^{(k)} = \left[ \int X(\tau) \exp(-jk\Omega\tau) d\tau \right] / T_0$.

Similar Fourier decompositions are assumed for the electric potentials ($\Phi_{EDL}$, $\Phi_{SCL}$, $\Phi_w$ and $\Phi_f$), net charge densities $\rho_i$ ($i$=1,2) and free surface charges $q$ in terms of their corresponding complex amplitudes $\Phi_{EDL}^{(k)}$, $\Phi_w^{(k)}$, $\Phi_f^{(k)}$ $\rho_i^{(k)}$ and $q^{(k)}$. Note that the component $k$=0 corresponds to a steady DC electric forcing.

The transformed problems resulting from Eqs. (2) together with the boundary conditions given by Eqs.(4) yield the complex amplitudes for net charge densities $\rho_i^{(k)}$. Substituting these two results into the right-hand sides of the transformed of Eqs. (1) and integrating twice with respect to $Y_1$ and $Y_2$, we obtain

$$\Phi_{EDL}^{(k)} = \frac{d\Phi_{EDL}^{(k)}}{dY_1}\bigg|_{Y_1=0} \left[ \left(1 - \frac{1}{\gamma_1^2}\right) Y_1 - \frac{1}{\gamma_1^3} e^{-\gamma_1 Y_1} \right] + A_1 \qquad (7a)$$

$$\Phi_{SCL}^{(k)} = z \frac{d\Phi_{SCL}^{(k)}}{dY_2}\bigg|_{Y_2=0} \left[ \left(1 - \frac{1}{\gamma_2^2}\right) Y_2 + \frac{1}{\gamma_2^3} e^{\gamma_2 Y_2} \right] + A_2 \qquad (7b)$$

with $\gamma_1^2 = 1 + jk\Omega$ and $\gamma_2^2 = 1 + t_w jk\Omega / t_f$. At the outer edges of the two inner regions, i.e., $Y_1 \to \infty$ and $Y_2 \to -\infty$, the solutions given by Eqs.(7) are going to be matched with the solutions outside the EDL and SCL to determine the unknown coefficients, and hence we have

$$\Phi_f^{(k)} = A_1 \quad \text{and} \quad \frac{d\Phi_f^{(k)}}{dy} = \frac{1}{\delta_1} \frac{d\Phi_{EDL}^{(K)}}{dY_1}\bigg|_{Y_1=0} \left(1 - \frac{1}{\gamma_1^2}\right) \quad \text{as } y \to 0 \qquad (8a)$$

$$\Phi_w^{(k)} = A_2 \quad \text{and} \quad \frac{d\Phi_w^{(k)}}{dy} = \frac{z}{\delta_2} \frac{d\Phi_{SCL}^{(K)}}{dY_2}\bigg|_{Y_2=0} \left(1 - \frac{1}{\gamma_2^2}\right) \quad \text{as } y \to 0 \qquad (8b)$$

The coefficient of the exponential term on the right-hand side of Eq. (7a) can be viewed as the complex amplitude for the effective induced zeta potential that is the potential drop across the EDL, i.e.,

$$\zeta_i^{(k)} = -\left(d\Phi_{EDL}^{(k)}/dY_1\bigg|_{Y_1=0}\right)/\gamma_1^3 = -\delta_1\left(d\Phi_f^{(k)}/dy\bigg|_{y=0}\right)/\left[\gamma_1\left(\gamma_1^2 - 1\right)\right] \qquad (9)$$

Making use of Eqs.(8), we eliminate $A_1$, $A_2$, $d\Phi_{EDL}^{(K)}/dY_1\big|_{Y_1=0}$ and $d\Phi_{SCL}^{(K)}/dY_2\big|_{Y_2=0}$ from the transformed boundary conditions Eqs. (3) to obtain the followings

$$\Phi_f^{(k)} - \Phi_w^{(k)} = \frac{d\Phi_f^{(k)}}{dy} \frac{\delta_1}{\gamma_1\left(\gamma_1^2 - 1\right)} + \frac{d\Phi_w^{(k)}}{dy} \frac{\delta_2}{\gamma_2\left(\gamma_2^2 - 1\right)} \qquad \text{at } y=0 \qquad (10a)$$

$$\delta_1 \frac{\gamma_1^2}{\left(\gamma_1^2 - 1\right)} \frac{d\Phi_f^{(k)}}{dy} - \beta\delta_2 \frac{\gamma_2^2}{\left(\gamma_2^2 - 1\right)} \frac{d\Phi_w^{(k)}}{dy} = 0 \qquad \text{at } y=0 \qquad (10b)$$

Eqs. (10) constitute the transformed version of the sought electrokinetic boundary conditions that directly connect complex amplitudes of two bulk potentials ($\Phi_w^{(k)}$ and $\Phi_f^{(k)}$) across the semiconductor-electrolyte interface. These derived electrokinetic boundary conditions are the key results of the present analysis. They are applicable to the AC induced-charge electrokinetic flow over solids of any dielectric constant and conductivity under an electric field with arbitrary wave forms. For all finite values of $\beta$ and $t_w/t_f$, the solution of the electrostatic problem consists of the simultaneous determination of the potentials $\Phi_f^{(k)}(\mathbf{r})$ and $\Phi_w^{(k)}(\mathbf{r})$ (wherein $\mathbf{r}$ denotes the position vector) which are harmonic (governed by Laplace equation) within the respective fluid and solid domains, and satisfy the boundary conditions (10) on the surface of the semiconducting solid (as well as the far-field conditions for $\Phi_f^{(k)}(\mathbf{r})$).

For conventional electrokinetic phenomena, solid walls are considered as perfect insulators, suggesting that both $\beta$ and $t_f/t_w$ are equal to zero. Then it can be obtained from Eqs. (10) that there is no induced zeta potential (since in this case there is no electric field inside the solid, and $\Phi_f^{(k)} - \Phi_w^{(k)}$ is the effective induced zeta potential drop across the EDL) and the bulk electrostatic potential inside the liquid domain satisfies the homogeneous Neumann condition, i.e., the electrically insulating condition ($d\Phi_f^{(k)}/dy = 0$).

For ideal dielectric objects under a DC electric field, the conductivity of the solids is zero ($t_f/t_w = 0$), there is no SCL effect inside the solid and the frequency of external electric field is zero ($\Omega = 0$), and Eqs.(10) reduce to

$$\Phi_w^{(k)} + \frac{\varepsilon_w \lambda_{D_1}}{\varepsilon_f a} \frac{d\Phi_w^{(k)}}{dy} = \Phi_f^{(k)} \text{ and } \frac{d\Phi_f^{(k)}}{dy} = 0 \qquad \text{at } y=0 \qquad (11)$$

which shows that the Robin-type and Neumann-type boundary conditions of the steady electrostatic problems for the solid wall and in the bulk liquid, respectively. The detailed derivation of Eq. (11) was provided in refs.[15](see their Eqs. (2) and (9)).

To gain further physical insight into the two boundary conditions given in Eqs. (10), we substitute Eq. (9) into Eqs.(10) to obtain the charging equation for the EDL and the SCL

$$\frac{d\Phi_f^{(k)}}{dy} = -\frac{\zeta_i^{(k)}}{\delta_1}(\gamma_1^2 - 1)\gamma_1 \qquad \text{at } y=0 \qquad (12a)$$

$$\frac{d\Phi_w^{(k)}}{dy} = -\frac{\Phi_w^{(k)} - \Phi_f^{(k)} - \zeta_i^{(k)}}{\delta_2}(\gamma_2^2 - 1)\gamma_2 \qquad \text{at } y=0 \qquad (12b)$$

In Eqs. (12), the left-hand sides represent the instantaneous Ohmic charging rates at the outer edges of the EDL and SCL, which are equal to the growth rates of their total induced charges shown on the right-hand sides of Eqs. (12). It is noted that $\Phi_w^{(k)} - \Phi_f^{(k)} - \zeta_i^{(k)}$ in Eq. (12b) denotes the potential drop across the SCL and is the counterpart of induced zeta potential $\zeta_i^{(k)}$ in the solid. If the EDL and SCL are to be considered as effective capacitors, we also can obtain frequency dependent capacitances (in dimensional form) for the EDL as $\varepsilon_0 \varepsilon_f \gamma_1 / \lambda_{D1}$ and the SCL as $\varepsilon_0 \varepsilon_w \gamma_2 / \lambda_{D2}$. Therefore, they provide a rigorous alternative to the widely used equivalent RC circuit models for the EDL in AC electrokinetic phenomena, where capacitance of EDL reads $\varepsilon_0 \varepsilon_f / \lambda_{D1}$ and is independent of frequency [1, 23, 25]. Also shown in the above conditions is the parameter $\beta = (\varepsilon_w \lambda_{D1})/(\varepsilon_f \lambda_{D2})$ which, based on the equivalent RC-circuit model, represents the ratio of the capacitance of the SCL $\varepsilon_0 \varepsilon_w / \lambda_{D2}$ and that of the EDL $\varepsilon_0 \varepsilon_f / \lambda_{D1}$. From this analogy it is anticipated that, when $\beta \gg 1$, the difference $\Phi_w^{(k)} - \Phi_f^{(k)}$, effectively representing the complex amplitude of induced zeta potential $\zeta_i^{(k)}$, becomes of comparable magnitude as $\Phi_f^{(k)}$. In the limit $\beta \to \infty$ (i.e. a perfectly polarizable solid) $\Phi_w^{(k)} = 0$ (also $d\Phi_w^{(k)}/dy = 0$ since electric field disappear inside the solid) and our result Eq.(12a) then precisely reduces to the macro-scale model of ref. [6] (see their Eqs. (7.48) and (7.50)).

## III. ACEK PHENOMENA AROUND A SEMICONDUCTIVE CYLINDER

## A. Mathematical formulations and solutions

In this section, we use a concrete example of ACEK phenomena around a semiconducting cylinder (see Fig.2) to demonstrate the applications of the effective boundary conditions derived above. The semiconducting cylinder with radius of R immersed in an electrolyte solution is simultaneously floating in an AC electric field of sinusoidal wave form, i.e., $E = \text{Re}\left[ E_0 \exp(j\Omega\tau) \right]$ and Re ( ) denotes the real part of a complex number. Then EDL inside the liquid domain and SCL inside the solid domain develop near semiconducting surface. It is already mentioned that complex amplitudes of potentials inside bulk electroneutral liquid and solid domains, $\Phi_f$ and $\Phi_w$, are all governed by Laplace equation. And the boundary conditions connecting these two domains follows from Eqs. (10). In polar coordinates, they can be reformulated as

$$\Phi_f - \Phi_w = \frac{\partial \Phi_f}{\partial r} \frac{\delta_1}{\gamma_1(\gamma_1^2 - 1)} + \frac{\partial \Phi_w}{\partial r} \frac{\delta_2}{\gamma_2(\gamma_2^2 - 1)} \text{ at } r=1 \qquad (13a)$$

$$\delta_1 \frac{\gamma_1^2}{(\gamma_1^2 - 1)} \frac{\partial \Phi_f}{\partial r} - \beta \delta_2 \frac{\gamma_2^2}{(\gamma_2^2 - 1)} \frac{\partial \Phi_w}{\partial r} = 0 \text{ at } r=1 \qquad (13b)$$

In addition, we also need a far field condition for $\Phi_f$

$$\Phi_f = -E_0 x = -E_0 r \cos\theta \text{ as } r \to \infty \qquad (14)$$

In Eqs.(13) and (14), potentials, electric field strength and radial coordinate are respectively normalized with respect to $\Phi_0, \Phi_0 / R$ and R. Referring to definitions of two electrokinetic parameters, $\delta_1$ and $\delta_2$, in section II, two electrokinetic parameters $\delta_1$ and $\delta_1$ in this case can be written as $\delta_1 = \lambda_{D1}/R$, $\delta_2 = \lambda_{D2}/R$. For the given sinusoidal AC electric field, $\gamma_1^2 = 1 + j\Omega$ and $\gamma_2^2 = 1 + jt_w\Omega/t_f$.

The electrostatic potentials satisfy Laplace equation in both bulk liquid and solid domains and the assumed solutions for complex amplitudes of the potential inside the bulk electrolyte domain, $\Phi_f$, and inside the bulk semiconductive cylinder, $\Phi_w$, are [26]

$$\Phi_f = -E_0 \cos\theta \left( r + \frac{A}{r} \right) \tag{15a}$$

$$\Phi_w = -BE_0 r \cos\theta \tag{15b}$$

Substituting Eqs.(15) into the Eqs.(13), two unknowns in Eqs.(14) then can be determined as

$$A = 1 - \frac{2\beta G_2 \gamma_2^3}{G_1 \gamma_1^3 (1+G_2) + \beta G_2 \gamma_2^3 (1+G_1)} \tag{16a}$$

$$B = \frac{2G_1 \gamma_1^3}{G_1 \gamma_1^3 (1+G_2) + \beta G_2 \gamma_2^3 (1+G_1)} \tag{16a}$$

where $G_1$ and $G_2$ are two complex groups related to the liquid domain and solid domain, respectively

$$G_1 = \frac{\delta_1}{\gamma_1(\gamma_1^2 - 1)} \tag{17a}$$

$$G_2 = \frac{\delta_2}{\gamma_2(\gamma_2^2 - 1)} \tag{17b}$$

Furthermore, the tangential electric field strength on the semiconducting surface reads

$$E_\theta = -\frac{1}{r}\frac{\partial \Phi_f}{\partial \theta} = -E_0 \sin\theta (1+A) \text{ at } r=1 \tag{18}$$

In this particular case, the complex amplitude of the induced zeta potential defined by Eq (9) can be formulated as

$$\zeta_i = -\frac{\partial \Phi_f}{\partial r} \frac{\delta_1}{\gamma_1(\gamma_1^2 - 1)} = G_1(1-A)E_0 \cos\theta \text{ at } r=1 \tag{19}$$

It is evident from Eq. (19) that the induced zeta potential is linearly proportional to the external electric field strength. For conducting cylinder with ideal polarizability under a DC forcing, namely $\beta \to \infty$ and $\Omega \to 0$, one can find out the induced zeta potential is

$$\zeta_i = 2E_0 \cos\theta \tag{20}$$

which is identical to the result given by Eq.(3.5) in ref. [6].

Utilizing *Helmholtz-Smoluchowski* equation, electroosmotic slip velocity on the surface of the semiconducting cylinder can be found out as

$$\mathbf{u}_s = U_0 E_0^2 \sin\theta \cos\theta \widehat{\boldsymbol{\theta}} \tag{21}$$

where $\widehat{\boldsymbol{\theta}}$ denote the unit vector along the azimuthal directions. And $U_0$ can be determined as

$$U_0 = \text{Re}\left[G_1(1-A)\exp(j\Omega\tau)\right]\text{Re}\left[(1+A)\exp(j\Omega\tau)\right] \tag{22}$$

Once the slip velocity $\mathbf{u}_s$ has been obtained from the above solution of the electrostatic problem, we proceed to resolve the flow field around the semiconducting cylinder. The flow of the bulk electroneutral fluid is governed by the dimensionless continuity and Stokes equations,

$$\nabla \cdot \mathbf{u} = 0 \text{ and } -\nabla p + \nabla^2 \mathbf{u} = 0 \tag{23}$$

respectively, which are subjected to the slip velocity boundary condition

$$\mathbf{u} = \mathbf{u}_s \text{ at } r = 1 \tag{24}$$

and the far field boundary condition

$$\mathbf{u} = 0 \text{ as } r \to \infty \tag{25}$$

where **u** and $p$ represent the fluid velocity vector and the pressure. These two are here normalized by $\varepsilon_0 \varepsilon_f \Phi_0^2 / (\mu R)$ and $\varepsilon_0 \varepsilon_f \Phi_0^2 / R^2$, respectively.

It is convenient to make use of the stream function formulation to solve the hydrodynamic problem. First, one can express radial ($u_r$) and azimuthal ($u_\theta$) velocity components in terms of the stream function $\psi$

$$u_r = \frac{1}{r}\frac{\partial \psi}{\partial \theta}, \quad u_\theta = -\frac{\partial \psi}{\partial r} \tag{26}$$

where he stream function is normalized with respect to $\varepsilon_0 \varepsilon_f \Phi_0^2 / \mu$.

Then after substituting Eq. (26) into Eqs. (23), the hydrodynamic problem originally governed by Eqs. (23) can be reduced to a biharmonic equation

$$\nabla^2 (\nabla^2 \psi) = 0 \tag{27}$$

where operator $\nabla^2$ in polar coordinate can be expressed as

$$\nabla^2 = \frac{1}{r}\frac{\partial}{\partial r}\left(r\frac{\partial}{\partial r}\right) + \frac{1}{r^2}\frac{\partial^2}{\partial \theta^2} \tag{28}$$

Boundary conditions given by Eqs.(24) and (25) can be transformed to

$$\psi = 0, \quad \frac{\partial \psi}{\partial r} = -U_0 E_0^2 \sin\theta \cos\theta \text{ at } r=1 \tag{29}$$

as well as the far-field boundary conditions

$$\frac{1}{r}\frac{\partial \psi}{\partial \theta} = 0, \quad -\frac{\partial \psi}{\partial r} = 0 \text{ as } r \to \infty \tag{30}$$

Squires and Bazant [6] derived the solutions for the fluid motion around a perfectly polarizable cylinder immersed in an electrolyte solution under a DC external electric field (see their Table 1). By analogy, we find complex amplitudes for stream function and corresponding velocity components of the fluid flow outside the semiconducting cylinder to be

$$\psi = \frac{(1-r^2)}{4r^2} U_0 E_0^2 \sin 2\theta \qquad (31a)$$

$$u_r = \frac{(1-r^2)}{2r^3} U_0 E_0^2 \cos 2\theta \qquad (31b)$$

$$u_\theta = \frac{1}{2r^3} U_0 E_0^2 \sin 2\theta \qquad (31c)$$

The flow field scales nonlinearly with respect to the external electric field strength. This feature differs from classic electrokinetic flows over insulating surfaces where the flow field is linearly proportional to the external electric field strength. Again, under the limit of a conducting cylinder with ideal polarizability under a DC forcing, Eqs.(31) are shown to be reduced to the results given in [6]

$$\psi = \frac{(1-r^2)}{r^2} E_0^2 \sin 2\theta \qquad (32a)$$

$$u_r = \frac{2(1-r^2)}{r^3} E_0^2 \cos 2\theta \qquad (32b)$$

$$u_\theta = \frac{2}{r^3} E_0^2 \sin 2\theta \qquad (32c)$$

## B. Results and discussion

To show the basic flow patterns at difference phases around the cylinder, we choose a special case of conductor with ideal polarizability ( $\beta \to \infty$ ) and the corresponding contours for stream function (normalized with respect to $E_0^2$) of difference phases are plotted in Fig.3. Since the stream function value is zero on the semiconducting surface, and then the stream function value at a point in the flow field gives the volumetric flow rate through a line connecting that point and the semiconducting surface. The higher the flow rate is, the stronger the flow is. Then magnitude of the stream function can be seen as a measure of the flow strength. It is evident that the flow is strongest at $\Omega\tau=0$ (Fig.3a). Basic flow pattern involves four vortices symmetric with respect to both *x* and *y* axes. These circulations are consequences of induced slip velocities all directing towards *x*=0 along the cylinder surface. Then as time evolves, the external field strength and the induced zeta potential all decrease, and so do the induced slip velocity on the cylinder surface and the flow strength, see Fig.3b. To phase $\Omega\tau=\pi/2$ in Fig.3c, although the external electric field decrease to zero, the tangential electric field strength on the semiconducting surface defined by Eq.(18) and the induced zeta potential given by Eq.(19) are not zero due to the essential phase lags between these two and the external field. Consequently, the liquid slip on the cylinder surface and thus the weak circulations still persist. After phase $\Omega\tau=\pi/2$, the external electric field reverse its direction and the local induced zeta potential on the cylinder surface also reverse its sign. Thus direction of the induced slip velocity (product of the tangential electric field strength and the induced zeta potential) on cylinder surface remain the same as that in the first half cycle ($\Omega\tau$ from 0 to $\pi/2$), and so does the direction of flow circulations. With the increasing magnitude of the electric field strength, the flow becomes intensified as shown in Fig.3d. Till phase $\Omega\tau=\pi$ (not shown here), the flow is intensified to be the same as $\Omega\tau=0$. It is also worth mentioning that the frequency of fluid oscillation is doubled to $2\Omega$ since both driving electric field and induced zeta potential oscillate at $\Omega$.

If the case of an ideally polarized cylinder under a DC forcing is taken as a reference, and then the solutions of flow field around a semiconducting cylinder can be found by multiplications of those given by Eqs. (32) and the scale factor of $U_0/4$. This factor is important since the dynamics of the oscillating flow around the semiconducting cylinder are included in this factor, such as the period, the amplitude and phase. Fig.4 characterizes the dependence of $U_0/4$ on the frequency of external fields ($\Omega$) and the free charge relaxation time ratio ($t_w/t_f$). It is clear that the amplitude of the flow oscillation increases with the decrease of both $\Omega$ and $t_w/t_f$. One also notes that the phase lag between the induced flow and the external electric field (phase information characterized by cos ($\Omega\tau$)) also reduces with the decrease of $\Omega$ and $t_w/t_f$. These features can be interrelated as follows: when $\Omega$ is low, there is sufficient time for free charge carriers to diffuse into EDL and SCL and to charge them up; as the decrease of $t_w/t_f$, the solid become more conductive and the free charge inside semiconducting cylinder responds quickly to form the SCL equivalent of a surface charge on the semiconducting surface.

## IV. SUMMARY

To conclude, we have derived effective electric boundary conditions for ACEK phenomena over semiconducting solids floating in AC electric fields with arbitrary wave forms. The general electric boundary conditions take into account the contributions from both the electrical polarization and the electrical conduction to induced zeta potentials. We have demonstrated that our general boundary conditions can recover several well-known limiting cases, such as the conventional electrokinetic phenomena with perfect insulating surfaces and the induced charge electrokinetic phenomena with perfectly polarizable surfaces. The effective electric boundary conditions also allow us to analyze ACEK phenomena with induced EDL and SCL effects while without need of resolving the detailed information of the EDL and SCL. A case study of the ACEK phenomena around a semiconducting cylinder is presented to show the applications of the derived boundary conditions. Our analyses show that four flow circulations are induced around the semiconducting cylinder. Furthermore, the intensity of these circulations can be modulated by adjusting the parameters associated with the EDL and SCL.

Considering the fact that the SCL is actually an EDL in the solid, the derived effective boundary conditions can be extended to describe the dynamic behavior of the interface between two immiscible electrolyte solutions (ITIES) under AC electric forcing. ITIESs are widely used for biomimetics, catalysis, surface cleaning, and assembly of nanoparticle arrays [27, 28]. Monroe et al. [29, 30] investigated the behavior of ITIES under the DC forcing and pointed out that two EDLs forming near the ITIES play essential roles. Another potential application of presented boundary conditions is to modify the classic dielectrophoresis theory of semiconducting particles in electrolyte solutions. As is known to all, EDL and SCL are not considered in conventional theory of dielectrophoresis. And continuities of the electric potential and electric displacement are used to derive the dielectrophoretic force and torque acting on a particle. Since most of electrolyte solutions and solid particles are usually semiconductive, EDL and SCL of finite thicknesses develop inside the solution and the particles, respectively. Under this case, obviously, the continuity of electric potential does not hold anymore (see Eq. (10a)) due to the charging of EDL and SCL. Then the local electric field around the particle is modified accordingly. Ultimately, the effective dipole moment and therefore dielectrophoretic force and torque on the particle are definitely modified by the dynamic charging of EDL and SCL. Although some searchers [22, 31-33] noticed this problem and already addressed the effect of EDL charging on the dielectrophoretic force, SCL effect was not considered until now. Furthermore, they did not provide simple formulae for dielectrophoretic force and torque on a spherical particle with EDL and SCL effects. With our derived effective boundary conditions, it can be expected that the modification of convention dielectrophoretic theory can be included in a modified Clausius−Mossoti factor. These two works are in progress and will be reported elsewhere.

## Acknowledgement

The authors gratefully acknowledge the financial support from the Ministry of Education of Singapore to CY (RG17/05) and the research assistantship from Nanyang Technological University to CLZ.

**Figure legends**

FIG.1 A schematic representation of the electrostatic problem in four sub-domains, namely, (i) the bulk fluid domain $\Phi_f$, (ii) the bulk leaky dielectric solid wall domain $\Phi_w$, (iii) the EDL domain $\Phi_{EDL}$ inside the liquid and (iv) the SCL domain $\Phi_{SCL}$ inside the solid. The dash lines inside the fluid and solid wall respectively represent the outer edges of the EDL and SCL where $\Phi_{EDL}$ matches $\Phi_f$ and $\Phi_{SCL}$ matches $\Phi_w$, respectively. And $\lambda_{D1}$ and $\lambda_{D2}$ measure the thicknesses of EDL and SCL, respectively.

FIG.2 A semiconducting cylinder immersed in an unbounded electrolyte solution is simultaneously floating in an external AC electric field of sinusoidal wave form. The external electric field *E* is applied along the *x* direction. Coordinates are normalized with respect to radius of the cylinder (R) and the electric field strength is normalized with respect to $\Phi_0/R$.

FIG.3 Contours for stream function at difference phases for the case of conducting cylinder with perfect polarizability ($\beta\to\infty$), (a) $\Omega\tau=0$, (b) $\Omega\tau=\pi/4$, (c) $\Omega\tau=\pi/2$ (d) $\Omega\tau=3\pi/4$. The arrowed lines are stream lines. In calculations, electrokinetic parameter $\delta_1=1/100$ and the frequency $\Omega=0.001$.

FIG.4 Variation of $U_0/4$ with the phase angle. (a) Dependence of $U_0/4$ on the frequency of the external electric field, $\Omega$, when $t_w/t_f=1.0$. (b) Dependence of $U_0/4$ on the free charge relaxation time ratio, $t_w/t_f$, when $\Omega = 0.01$. $\beta=1$, $\delta_1 = \delta_2 = 1/100$ are chosen for all the calculations.

**FIG.1**

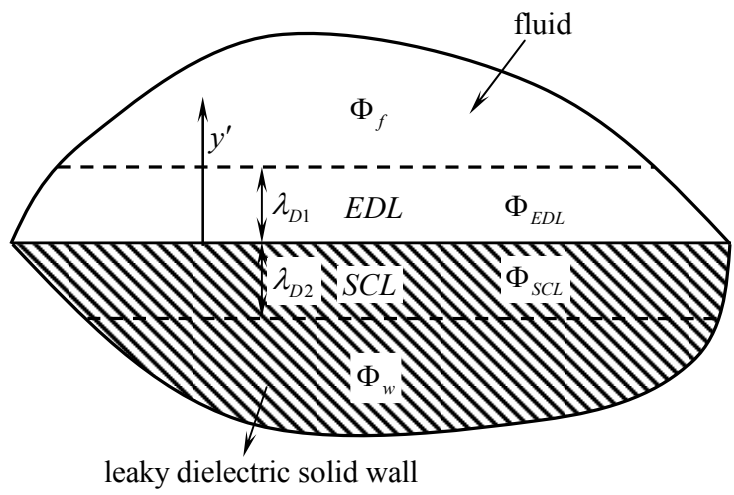

leaky dielectric solid wall

**FIG.2**

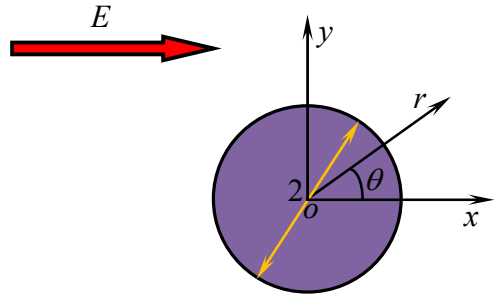

**FIG.3**

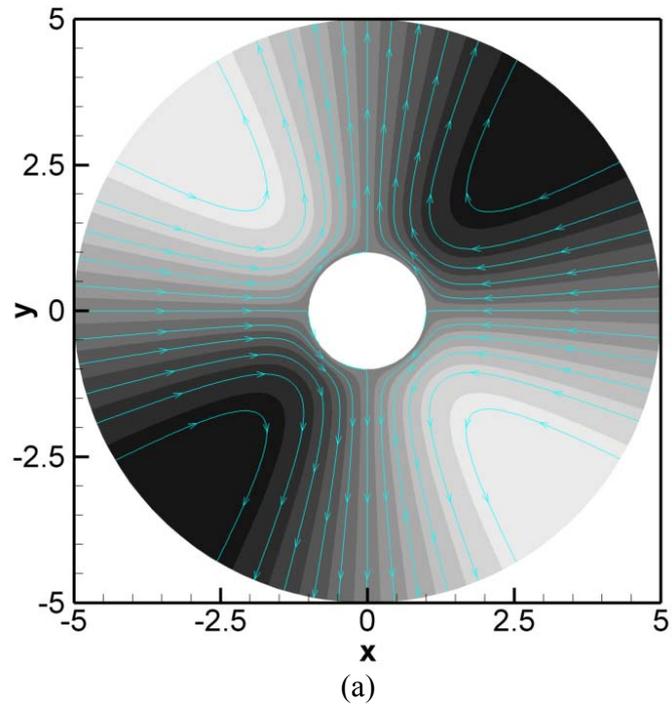

(a)

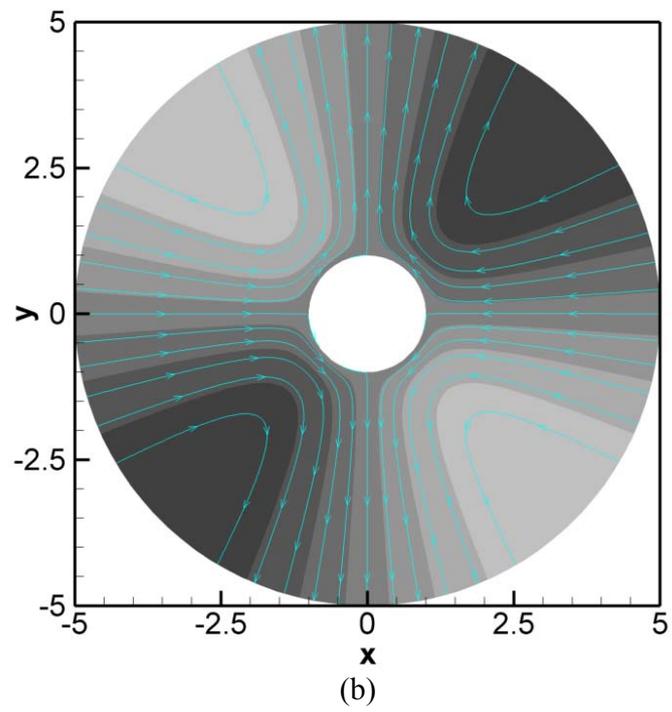

(b)

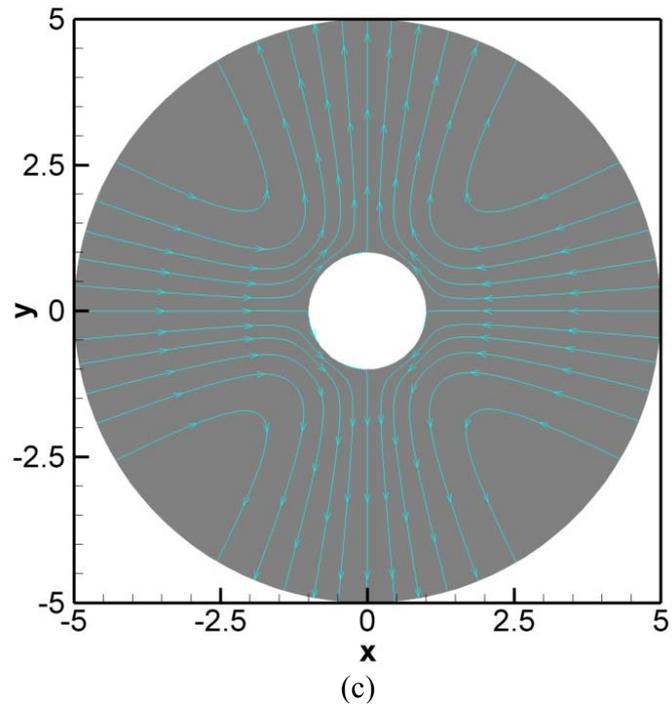

(c)

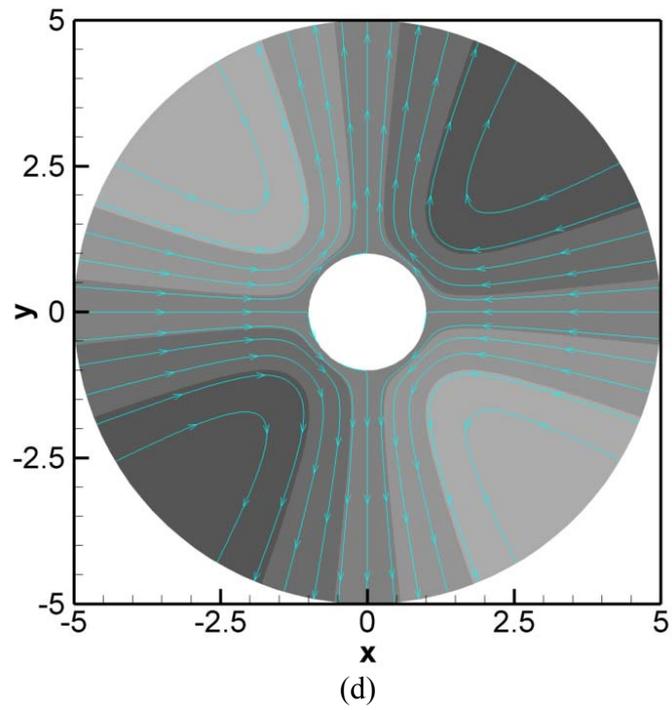

(d)

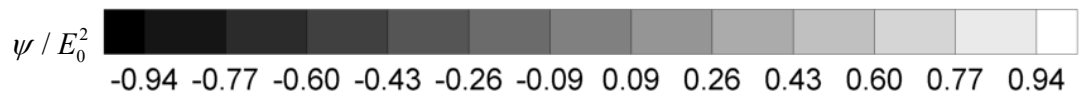
28

**FIG.4**

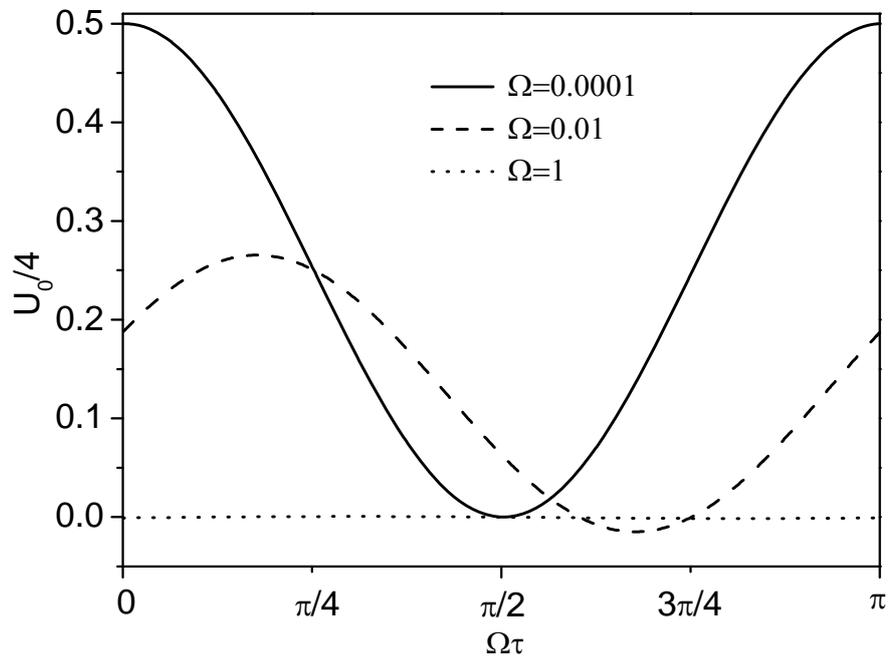

(a)

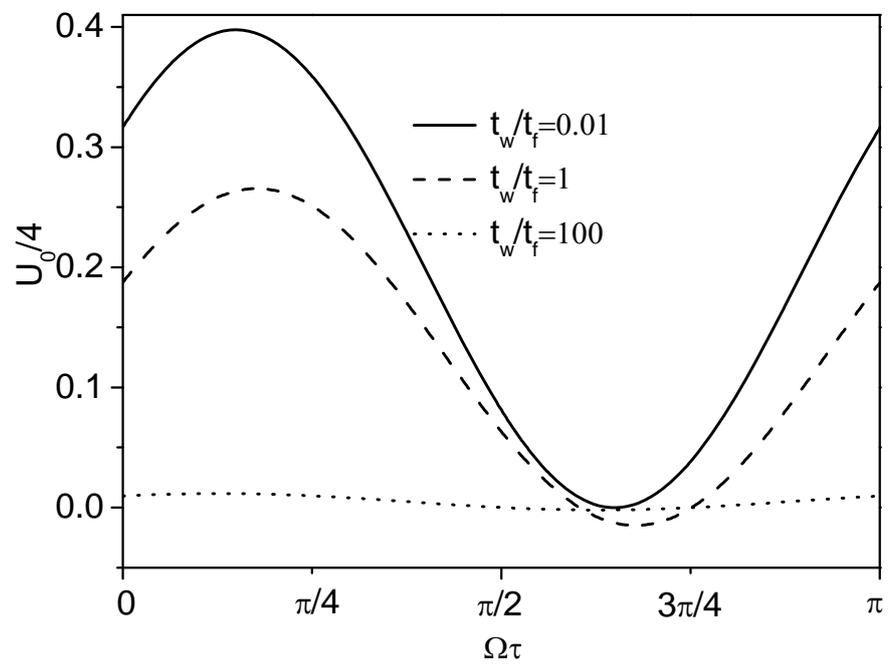

(b)